\documentclass[epsfig,twocolumn,showpacs,preprintnumbers,amsmath,amssymb]{revtex4}

\usepackage{graphicx}
\usepackage{dcolumn}
\usepackage{bm}

\begin{document}


\title{Effect of gap suppression by superfluid
current on nonlinear microwave response of d-wave superconductors}
\author{E.J. Nicol}
\email{nicol@physics.uoguelph.ca}
\affiliation{Department of Physics, University of Guelph,
Guelph, Ontario, N1G 2W1, Canada}
\author{J.P. Carbotte}
\email{carbotte@mcmaster.ca}
\affiliation{Department of Physics and Astronomy, McMaster University,
Hamilton, Ontario, L8S 4M1, Canada}
\date{\today}

\begin{abstract}
Recently several works have focused on the intrinsic nonlinear
current  in passive microwave filters as a tool for identifying
d-wave order parameter symmetry in the high $T_c$ cuprates. 
Evidence has been found for d-wave pairing in
YBCO and further work has ensued. Most of the theoretical
work has been limited to  low temperature because it has not included
the effect of the superfluid current on the energy gap. We find that
this effect leads to 
important corrections above $T\sim 0.2T_c$, while leaving the
$1/T$ low temperature behavior intact.  A twofold increase in the nonlinear
coefficient at temperatures of order $\sim 0.75 T_c$ is found, and
as $T\to T_c$ the nonlinearity comes entirely from the effects of the
superfluid current on the gap. Impurity scattering has been 
included and, in addition, signatures for the case of d+s-wave 
are presented.
\end{abstract}
\pacs{74.20.Rp,74.25.Nf,74.72.Bk}

\maketitle

\section{Introduction}

Interest in generating novel methods for probing the order parameter
symmetry in superconductors has driven the development of new
techniques and experimental configurations, such as angle-resolved
electron tunneling\cite{dswave} or angle-resolved magnetic field
dependence of specific heat\cite{specheat}, which can provide unambiguous
signatures of order parameter nodes and their location. One such
proposal is associated with the measurement of the intrinsic nonlinear
current in passive  microwave filters. Such devices give rise to
to third order intermodulation effects which, while detrimental for
practical applications in superconducting communication filter technology,
make these devices ideal for examining issues of order parameter symmetry.
Excellent progress in the field of high temperature superconductivity
has been made in this area.

Initially, Yip and Sauls\cite{yip,xu} proposed the examination of the
nonlinear Meissner effect (NLME) for evidence of d-wave gap symmetry in
the cuprates. Their predictions were not confirmed by experiments at the
time\cite{saulsexpt}, however, Dahm and Scalapino (DS)\cite{dahmdwave,dahmapj}
 proposed to examine a related
quantity: the intermodulation current or distortion (IMD) which arises
from the nonlinear inductance resulting from a quadratic 
dependence
of the penetration depth on the superfluid current. In their work,
a signature of the d-wave gap would be found in an upturn in the temperature
dependence of the nonlinear coefficient at low temperatures, in contrast to
exponential decay for an s-wave gap. Such evidence of an upturn 
has been found in YBCO films by several groups\cite{benz,oates,leong}
with excellent agreement with the DS theory. This success has led to further
calculations to examine the issue of nonlocal effects for this quantity\cite{agassi}
(thought to be the possible reason for not seeing the NLME proposed by
Yip and Sauls\cite{mrli}) and the d-wave signature remains robust. 
More recently,
measurements of the nonlinear current have been made near $T_c$ on a range of
different films and the data has been analyzed in terms of the DS approach
with impurity scattering being used to explain
variations between the films\cite{andersen}.
 However, in this work and others\cite{salkola,dahmmg} which extrapolate 
the DS theory to high temperature, the effect of the superfluid current
on the gap has been neglected as an approximation. We show here that
inclusion of this effect has significant impact on the nonlinear coefficient
for $T \gtrsim 0.2 T_c$, requiring a reanalysis of previous results. Indeed,
when the current-dependence of the gap is considered, the
nonlinear coefficient shows a twofold increase at $T\sim 0.75 T_c$ over
the value calculated using the approximation of a current-independent gap.

Finally, we note that other theoretical works 
have examined the intrinsic nonlinear current
for two-band superconductors and MgB$_2$\cite{dahmmg,nicolmg} and for
one-band s-wave superconductors\cite{xu,nicolmg}. Effects of the
superfluid on the gap have been included in Refs.~\cite{xu,nicolmg}.

Our paper is structured as follows: in the next section, we summarize
our theoretical approach which allows for both the inclusion of the current
dependence in the gap and impurity scattering (from unitary to Born limit).
Strong electron-boson coupling effects are also available in this
formalism. In Section~III, we discuss the results illustrating the corrections
to the simplified theory at finite temperature and we revisit the
experimental situation, including the issue of impurities. We
end by providing predictions for an admixture of d- and s-wave
symmetry as has been recently suggested by angle-resolved electron tunneling
experiments on YBCO\cite{dswave}. 
We form our conclusions briefly in Section~IV.

\section{Theory}

In this work, we evaluate the full current from the standard expression
given for the imaginary axis Matsubara representation and modified for a
d-wave order parameter $\Delta(\theta)=\Delta\cos(2\theta)$ in 
two-dimensions:\cite{maki,xu,nicolmg,nicoljc,mgjcnicol}
\begin{eqnarray}
j_{s}(q_s,\alpha)&=&\frac{2en}{mv_{F}}\pi T\sum_{n=-\infty}^{+\infty}
\int_0^{2\pi} \frac{d\theta}{2\pi}\nonumber\\
&&\times\frac{i[\tilde\omega _n-is\cos(\theta-\alpha)]\cos(\theta-\alpha)}
{\sqrt{[\tilde\omega _n-is\cos(\theta-\alpha)]^2+
\tilde\Delta^2 _n\cos^2(2\theta)}}.
\label{eq:js}
\end{eqnarray}
This expression contains both the condensate current and the quasiparticle
current due to excitations.\cite{comment} 
The angle $\alpha$ measures the direction of the
current with respect to the order parameter antinode, with $\alpha=0$
indicating the current in the antinodal direction and $\alpha=\pi/4$ for
the nodal direction. The other notation is standard with
$e$ the electric charge, $m$ the electron mass, $T$ the
temperature,
$n$ the electron density and
$v_{F}$ the Fermi velocity.
The superfluid momentum $q_s$ enters through $s=v_{F}q_s$, and, importantly for
the results of this paper, it also enters the equations for the Matsubara gaps and
renormalized frequencies and, as a result, the current will decay the gap. The
equations for the Matsubara gaps $\tilde\Delta_n=Z_n\Delta_n$
and renormalized frequencies $\tilde\omega_n=\omega_nZ_n$ modified for d-wave
symmetry are:
\begin{eqnarray}
\tilde\Delta_n &=& \pi T\sum_{m=-\infty}^{+\infty}\lambda(m-n)
\int_0^{2\pi} \frac{d\theta}{2\pi}\nonumber\\
&&\times\frac{\tilde\Delta_m\cos^2(2\theta)}
{\sqrt{[\tilde\omega_m-is\cos(\theta-\alpha)]^2+\tilde\Delta_m^2\cos^2(2\theta)}}
\label{eq:Del}
\end{eqnarray}
and
\begin{equation}
\tilde\omega_n = \omega_n+g\pi T\sum_{m=-\infty}^{+\infty}
\lambda(m-n)\Omega_m
+ \pi \Gamma^+\frac{\Omega_n}{c^2+\Omega_n^2},
\label{eq:Z}
\end{equation}
with
\begin{equation}
\Omega_n=\int_0^{2\pi}\frac{d\theta}{2\pi}
\frac{\tilde\omega_n-is\cos(\theta-\alpha)}
{\sqrt{[\tilde\omega_n-is\cos(\theta-\alpha)]^2+\tilde\Delta_n^2\cos^2(2\theta)}},
\label{eq:Omeg}
\end{equation}
where  the Matsubara frequencies are
$\omega_n=\pi T(2n-1)$, for integer $n$, and
we have included the possibility of impurity scattering
from the unitary ($c=0$) to Born ($c\to \infty$) limit via the last
term in Eq.~(\ref{eq:Z}) which requires self-consistency through $\Omega_n$.
Here, $\Gamma^+$ is proportional to 
the impurity scattering rate and $c$ is related to the
the s-wave scattering phase shift.\cite{andersen}
Note that an impurity term does not appear in Eq.~(\ref{eq:Del}) in
d-wave as it does in s-wave. This is because in d-wave it averages to zero.
Finally, in general, the kernel of these equations would normally be based on a
momentum-dependent electron-boson spectrum. To mimic this unknown spectrum
in an approximate manner,
a parameter $g$ is introduced to represent that the interaction in the
$\omega$-channel could be different from that in the $\Delta$-channel
in the case of a momentum-dependent interaction that would give rise
to a d-wave order parameter symmetry. Likewise, there is no
$\cos(2\theta)$ factor in the numerator of the $\omega$-channel
 as there is in the $\Delta$-channel
reflecting that the interaction in the renormalization channel is taken
to be isotropic to first order.\cite{arberg,asymp} Thus, the electron-boson spectral function,
which we denote by $\alpha^2F(\Omega)$, enters
$\lambda(n-m)$ as follows:
\begin{equation}
\lambda(m-n)\equiv 2\int^\infty_0\frac{\Omega\alpha^2F(\Omega)}{
\Omega^2+(\omega_n-\omega_m)^2}d\Omega.
\label{eq:lambda}
\end{equation}
Here, to take the limit of these equations to give the standard BCS
result for d-wave in our numerical
evaluation, we take the electron-boson spectrum to be a delta function
at high frequency.  Likewise, we exclude renormalization effects 
which are not based on impurities by taking $g$ to be zero.
This gives the BCS gap ratio in d-wave to
be $2\Delta_0/kT_c=4.28$. When we wish to
consider strong-coupling effects corresponding to $2\Delta_0/kT_c=5$,
for example, then we take $g$ to be a finite value 
and use a delta function at lower frequency for the boson spectrum\cite{asymp}.

To extract the nonlinear coefficient that is relevant to the
passive microwave filters and hence can be used as a sensitive 
probe of order parameter symmetry, we can assume that $j_s$
can be expanded to third order for small $q_s$ as:
\begin{equation}
j_s=j_0\biggl[\frac{n_s(T)}{n}\biggl(\frac{q_sv_F}{\Delta_0}\biggr)
-\beta(T)\biggl(\frac{q_sv_F}{\Delta_0}\biggr)^3\biggr],
\label{eq:jexpand}
\end{equation}
 where
 $j_0=ne\Delta_0/(mv_F)$. Note for strong coupling, we replace
we replace the $q_sv_F$ in this formula by $q_sv_F/(1+\lambda)$,
where $\lambda$ is the mass renormalization parameter\cite{nicolmg}. 
From this
Dahm and Scalapino define\cite{dahmdwave}
\begin{equation}
b(T)\equiv \frac{\beta(T)}{[n_s(T)/n]^3},
\label{eq:dougb}
\end{equation}
the square of which is related to the third order intermodulation power in 
microwave filters.
Hence, measuring the intermodulation power provides a measure of the
nonlinear coefficient. As we calculate the full $q_s$ dependence of
$j_s$, using Eqs.~(\ref{eq:js})-(\ref{eq:lambda})
 with no approximation of taking the gap to be independent
of $q_s$, as was done in previous works, it is easiest to
extract these quantities directly from our numerical data.  
To do this we form  the quantity of $j_s/q_s$ versus $q_s^2$
which is a straight line at low $q_s$
and from this we obtain the superfluid density from the intercept
and the nonlinear coefficient from the slope. By this method,
we have confirmed previous results at low temperature and can
proceed to examine the issue of higher temperatures where the
current reduces the gap even at low $q_s$. We have also demonstrated this
method for one-band and two-band s-wave superconductors.\cite{nicolmg}

\section{Results}

\begin{figure}[ht]
\begin{picture}(250,200)
\leavevmode\centering\includegraphics{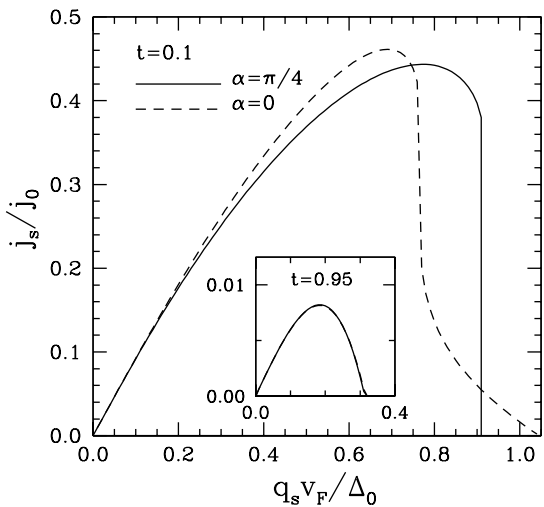}
\end{picture}
\caption{The normalized current $j_s/j_0$ as a function of $q_sv_F/\Delta_0$,
where $j_0=ne\Delta_0/(mv_F)$ and
$\Delta_0$ is the energy gap at $T=0$. Shown are the low temperature
BCS curves for two directions: the current in the antinodal direction with
$\alpha=0$ 
(dashed line) and the nodal direction with
$\alpha=\pi/4$ (solid), given for a reduced temperature
$t=T/T_c=0.1$. The inset shows that the curves overlap
for $T$ near $T_c$ (in this case, $t=0.95$).
}
\label{fig1}
\end{figure}

In Fig.~\ref{fig1}, we show the current as a function
of $q_s$ in the two major directions,
along the node and antinode, at both low and high temperatures.
This was done using the equations above and illustrates that we can
reproduce correctly the $T=0$ results in the literature\cite{ting,hykee} 
and that
we can also evaluate the current at high temperature in this formalism.
The Matsubara formalism is also ideally suited for including 
impurity scattering. There are few points to note in this figure. The
$q_s$ in the order parameter is essential to obtain these
curves and it is the decay of the order parameter by the superfluid
current that causes the current $j_s$ to drop dramatically beyond
the peak (otherwise, if $\Delta(q_s=0)$ is used, the curves decay
slowly to zero as $q_s \to \infty$,
for example, as $1/q_s$ for $T=0$ BCS s-wave). 
Furthermore, the effect of the $q_s$ in the
gap becomes even more important at low $q_s$ when the temperature is
approaching $T_c$. Finally, for high temperatures near $T_c$,
the current is fairly independent of the angle $\alpha$.

\begin{figure}[ht]
\begin{picture}(250,200)
\leavevmode\centering\includegraphics{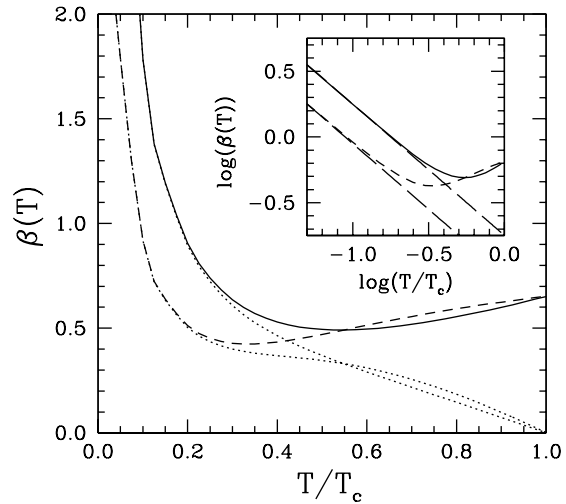}
\end{picture}
\caption{The nonlinear coefficient $\beta(T)$ as a function of
$T/T_c$, shown for two directions: $\alpha=\pi/4$ (solid line) and
$\alpha=0$ (short-dashed). The dotted curves are for the same two
directions but with the approximation of neglecting the $q_s$ dependence
in the gap, {\it i.e.} $\Delta(q_s=0)$. The inset illustrates via a log-log
plot that the low temperature behavior varies as $\Delta_0/24T$
for $\alpha=0$ and $\Delta_0/12T$ for $\alpha=\pi/4$ (these
expressions are shown as the long-dashed lines in both cases).
}
\label{fig2}
\end{figure}

\begin{figure}[ht]
\begin{picture}(250,200)
\leavevmode\centering\includegraphics{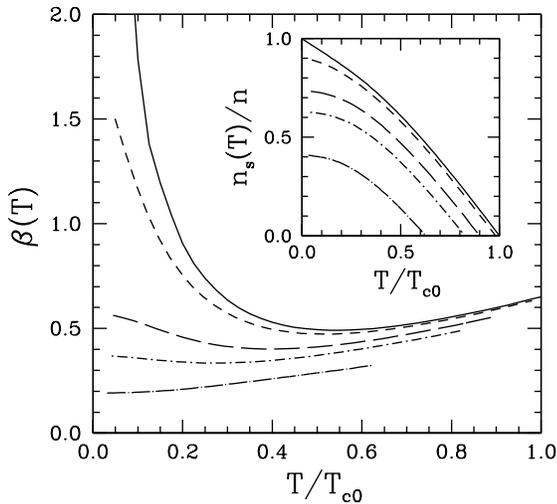}
\end{picture}
\caption{The nonlinear coefficient $\beta(T)$ versus $T/T_{c0}$
for varying impurity scattering in the unitary limit ($c=0$)
and  with
$\alpha=\pi/4$.
The inset shows the corresponding curves for the superfluid density
$n_s(T)/n$. Curves are given for: the pure limit (solid),
$\Gamma^+/T_{c0}=0.0082$ (short-dashed), 0.0404 (long-dashed),
0.0697 (dot-short-dashed), 0.1432 (dot-long-dashed).
}
\label{fig3}
\end{figure}

This latter feature is seen more clearly in Fig.~\ref{fig2}
where we show the nonlinear coefficient $\beta(T)$ for d-wave in 
the clean limit as a function of temperature for the two directions
just discussed (solid and short-dashed curves). 
Once again, it is seen that while the two
curves are different at low $T$, as $T\to T_c$, the anisotropy
is reduced and disappears at $T_c$. Indeed, we can obtain an analytical
value for $\beta(T)$ at $T=T_c$ which is 0.651 and this is confirmed
by the numerics shown in Fig.~\ref{fig2}. It is also important
to note that the same procedure which gives rise to these curves
also provides the normalized superfluid
density which is shown as the solid curve in the inset of Fig.~\ref{fig3}.
This superfluid density curve is independent of the direction
of the current as expected as it arises from the $q_s\to 0$ limit
and this curve is exactly the same as that which is
calculated by standard formulas for the penetration depth.
Here it is extracted from our $j_s$ versus $q_s$ curve as 
explained in the theory section, verifying the accuracy of our method.

Also shown in Fig.~\ref{fig2} by dotted line type are the curves
for $\beta(T)$ for $\alpha=0$ and $\pi/4$ when the approximation
of $\Delta(q_s=0)$ is taken ({\it i.e.}, the $s$ is set equal to
zero  in Eqs.~(\ref{eq:Del})-(\ref{eq:Omeg}) so that the gap is not modified
by the superfluid  current). This is a central point of our paper,
that while this approximation works well at low temperatures, one sees
from the comparison of the dotted curves with their respective short-dashed
and solid ones, that this approximation breaks down for $T\gtrsim0.2T_c$
and produces significant deviations as $T\to T_c$. Indeed, from physical
grounds one does not expect the nonlinear current to go to zero as 
$T\to T_c$, but rather to increase. From the point of view of device
applications which would typically operate at about $0.5T_c$ or higher,
this effect of decaying the gap by the superfluid can introduce a factor
of 1.5-2 increase in the nonlinear response of the device. Finally,
we confirm in Fig.~\ref{fig2}  that the original use of this approximation
for low temperatures is robust and the $1/T$ signature of the
d-wave gap discussed by Dahm and Scalapino\cite{dahmdwave,dahmapj} 
remains intact. We find by
our procedure, the same result as determined by Dahm and Scalapino
analytically, that $\beta(T,\alpha=0)\simeq\Delta_0/24T$ and
$\beta(T,\alpha=\pi/4)\simeq\Delta_0/12T$ for $T\to 0$, which is
illustrated by the log-log plot in the insert of Fig.~\ref{fig2}.
 
Turning to the case of impurities which were discussed previously by
Dahm and Scalapino\cite{dahmdwave} and Andersen {\it et al.}\cite{andersen},
we comment on the modifications that occur due to the current in the
gap. Again, the results of Dahm and Scalapino which are given for 
$T<0.2T_c$ remain robust, however, the results of Andersen {\it et al.}
which focus on high temperature are necessarily modified when the
approximation of a $q_s$-independent
order parameter is removed. This is shown in Fig.~\ref{fig3}, where
we show both $\beta(T)$ and $n_s(T)/n$ for varying impurity content
(here we use the unitary limit for illustration, as was done in Ref.~\cite{andersen}). The $\beta(T)$ curves are shown only for $\alpha=\pi/4$. Once
again, for $T\gtrsim 0.2T_c$, deviations occur in the manner discussed before.
The significant feature is that the impurity scattering 
reduces the $T_c$ relative to the pure 
case which has $T_{c0}$ and hence for fixed temperature of $T\sim 0.75 T_{c0}$,
for example, with varying impurity content, the decay of the gap by the
superfluid current has an even greater effect for increasing impurity content
and so it must not be neglected.

\begin{figure}[ht]
\begin{picture}(250,200)
\leavevmode\centering\includegraphics{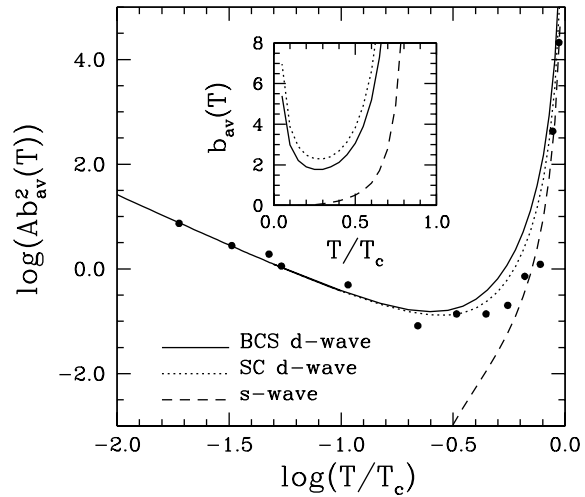}
\end{picture}
\caption{Plot of log($Ab^2_{av}(T)$) versus log($T/T_c$) for
s-wave (dashed curve), BCS d-wave with $2\Delta_0/kT_c=4.28$ (solid),
and strong-coupling (SC) d-wave with $2\Delta_0/kT_c=5$ (dotted).
The curves are compared with the experimental data (solid dots)
taken from Ref.~\cite{oates}. The inset shows the same curves for
$b(T)$ versus $T/T_c$.
}
\label{fig4}
\end{figure}

We now turn to a re-examination of the comparison of theory with some of
the data that exists in the literature, maintaining the same analysis 
that was done previously. First, we consider the clean limit YBCO
data presented by Oates and coworkers\cite{oates}. In Fig~\ref{fig4},
we reproduce this data for the normalized IMD along with our
calculations for BCS d-wave, where $2\Delta_0=4.28kT_c$. The
IMD power has been shown to be related to $b^2(T)$, {\it i.e.}
\begin{equation}
P_{\rm IMD}\propto b^2(T)
\end{equation}
and hence we plot $b^2(T)$ in the figure and use an adjustable factor
$A$ to match the data with the theory as was done by Oates {\it et al.}.
Likewise, we have averaged the two directions for $\alpha=0$ and $\alpha=\pi/4$
after the manner of Oates {\it et al.} as their measurement averages over
all directions. The low temperature part of this log-log plot does not
change from previous d-wave calculations but the high temperature part
is increased and the agreement with the data is not as good in this
regime (we have chosen to fit the low temperature part of the curve to the
data). However, compared to the s-wave calculation, the evidence in
support of d-wave remains striking. Oates {\it et al.} obtained
a better fit with the d-wave theory because Dahm and Scalapino used a larger
gap ratio in their BCS theory which is supported by other
experiments. To illustrate this within our formalism, we can do a strong-coupling 
calculation where we take $g=1$ in Eq.~\ref{eq:Z} and move the electron-boson
spectral function to lower frequency. In this manner, we can obtain
$2\Delta_0/kT_c=5$ in d-wave corresponding to a mass renormalization
parameter $\lambda=5$. In this case, the result (shown as the
dotted curve and plotted with a new $A$ to fit the low $T$ region) does
indeed move toward a better fit with the data over the full temperature
range. As a technical note, to obtain a gap ratio of $2\Delta_0=6kT_c$
in this Eliashberg formalism, we would have to take $\lambda$ to extremely
large values
and we would then be in the asymptotic regime which limits the value of
the gap ratio to go no higher than about 6.5.\cite{asymp} A much better fit is
obtained with a gap ratio of about 6 but in our model the $\lambda$ is
unphysical. To overcome this, in order to produce higher
gap ratios with more physical values of $\lambda$, 
would require a spin-fluctuation
calculation of the type given in Ref.~\cite{spin}, which 
would include feedback on the electron-boson
spectral density itself, which suppresses it at low $\omega$
and introduces coupling to a new resonant mode that grows in 
amplitude as $T$ is reduced.
This is beyond the scope of this work.
Other work which has discussed this data has been that
of Agassi and Oates\cite{agassi} where they considered nonlocal
effects, however, the calculation is a BCS one with no inclusion of
the effect of 
the superfluid current on the gap and so it is difficult to say how
both this and strong coupling might modify their results.

Finally in Fig.~\ref{fig4}, in the inset, we give for
reference the unscaled $b(T)$ versus $T$ curves. The $1/T$ divergence
is clear in comparison with the exponential decay of the s-wave case
and the strong-coupling calculation is above the BCS one as is expected
from previous work on s-wave superconductors\cite{nicolmg}.

\begin{figure}[ht]
\begin{picture}(250,200)
\leavevmode\centering\includegraphics{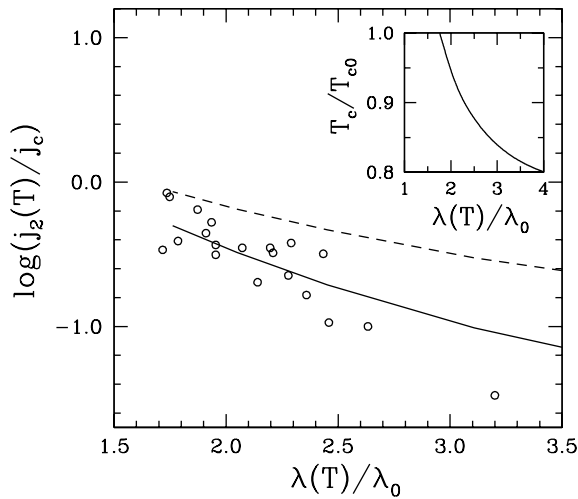}
\end{picture}
\caption{Plot of log($j_2(T)/j_c$) versus $\lambda(T)/\lambda_0$ for
$T=0.75T_{c0}$.  Andersen {\it et al.}\cite{andersen}
have defined a quantity $(j_2/j_c)^2$ which is equal to
our $1/b(T)$. The data (open circles) is taken from Ref.~\cite{andersen}.
The solid curve is for BCS d=wave with $\Delta(q_s)$ and the dashed is for the
case where the approximation $\Delta(q_s=0)$ has been used. The inset shows
the reduction in $T_c$ from the pure value of $T_{c0}$ that is implied by
the values of $\lambda(T)$ at $T=0.75T_{c0}$
normalized to the pure value for $T=0$, $\lambda_0$.
}
\label{fig5}
\end{figure}

We revisit the issue of impurities in Fig.~\ref{fig5}. This figure is
based on a similar one presented by Andersen {\it et al.}\cite{andersen}.
Several films were examined at $75$K and the nonlinear coefficient was
found to vary from sample to sample, as did also the penetration depth.
The analysis in the paper used the DS theory with impurity scattering
to argue that the data could be understood by assuming varying impurity
content from one film to another. We re-examine this issue because at this
high temperature the pairbreaking effect of the superfluid current on
the gap should cause significant changes (our Fig.~\ref{fig3} should be
contrasted with Fig.~3 of Ref.~\cite{andersen}) and this was not included in the
original analysis. In Fig.~\ref{fig5}, we show our calculation
at $T=0.75T_{c0}$  for the variation of the nonlinear coefficient 
with respect to the variation of the penetration depth when unitary
scattering is included (like Andersen {\it et al.}, we have checked Born
scattering and find
there is essentially no difference in the result presented here). We have
also taken $\alpha=\pi/4$ as at this temperature the anisotropy is almost
completely gone, which we have checked. 
It should be noted that, once again, there is an adjustable
parameter $j_c$ which can be used to scale the theory to overlap with the
data. Andersen {\it et al.} kept the theory fixed and used a value of $j_c$
to adjust the data. Here we choose to keep the scaled data as it 
was presented in the original paper and so we adjust our theory
by a scale factor to overlap with the data (if we had adjusted the
data instead, this would have increased the $j_c$ value used for scaling
the data by a factor of about 1.35).
Doing so, we find with the solid curve that there is reasonable
agreement between theory and data. However, we note, referring the 
reader to the inset of
the figure, that the range of variation in penetration depth, if explained
via impurity scattering, would imply a reduction in $T_c$ of about
15-20\% from the pure case\cite{quinlan} and yet the experimental data indicates
a $T_c$ variation of only about 2\%. This discrepancy was not noted in 
Ref.~\cite{andersen} and it does pose a problem for this interpretation.
However, continuing with this explanation, we have compared our result
with the case of taking $\Delta(q_s=0)$, shown as the dashed curve and
using the same $j_c$ scale factor, and we find that while some
deviation may be corrected by choosing a different $j_c$, which would shift the
curve, the slope is also different.
While the interpretation of the data via impurity scattering may be open
to question due to the large variation in $T_c$ required,
more experimental data may help to reduce the scatter and provide further
tests of the theory.

\begin{figure}[ht]
\begin{picture}(250,200)
\leavevmode\centering\includegraphics{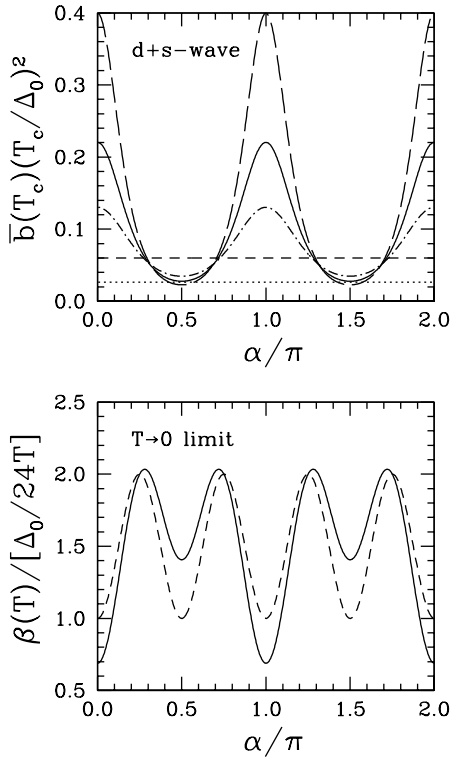}
\end{picture}
\vskip 70pt
\caption{The effect of a d$+$s-wave gap on the anisotropy near
$T_c$ as measured via $\bar b(T_c)$ (upper frame) and at
low temperature as measured by $\beta(T)$ normalized to $\Delta_0/24T$
(lower frame). The anisotropy is plotted as a function of the current 
direction relative to the antinode, $\alpha$. Curves are shown for
s-wave (dotted), d-wave (short-dashed) and varying percentage of
s-wave: 10\% (dot-dashed), 15\% (solid), 20\% (long-dashed).
}
\label{fig6}
\end{figure}

Recently, a novel probe of angle-resolved electron tunneling was used 
to determine the order parameter symmetry in YBCO and this work has 
suggested that the order parameter is not pure d-wave but an admixture of
d$+$s with about 15\% of s-wave. This is not the first time that it has
been suggested that YBCO may have a small s-wave component and it is 
important to assess the consequences for the intrinsic nonlinear current
and whether there would be any signatures unique to a d$+$s-wave order
parameter. As a result, we now briefly consider such a symmetry,
where we assume that the s-wave component is small compared with the
d-wave part, such that nodes still exist but they are shifted from 
$\pi/4$. Taking $\Delta(\theta)=\Delta[a\cos(2\theta)+c]$, we evaluate
analytically the main features expected in the limits of $T\to T_c$ and
$T\to 0$. As $T\to T_c$, we find that we can define a quantity related to
$b(T)$:
\begin{equation}
\bar b(T_c) \equiv \lim_{T\to T_c} b(T)\biggl(1-\frac{T}{T_c}\biggr)^3,
\end{equation}
which gives $(T_c/\Delta_0)^2\bar b(T_c)=0.0599$ for d-wave
and and 0.0266 for two-dimensional s-wave.
In both cases, this quantity is isotropic with
respect to the direction of the current, however, for d$+$s-wave, we have
\begin{equation}
\bar b(T_c)\biggl(\frac{T_c}{\Delta_0}\biggr)^2 = \frac{7\zeta(3)}{64\pi^2}
\frac{\gamma_2^2\gamma_3}{\gamma_1^3(\gamma_1-\gamma_3)^2},
\end{equation}
where
\begin{eqnarray}
\gamma_1 &=&\int_0^{2\pi} \frac{d\theta}{2\pi}
\Delta^2(\theta),\\
\gamma_2 &=&\int_0^{2\pi} \frac{d\theta}{2\pi}
\Delta^4(\theta),\\
\gamma_3 &=&\int_0^{2\pi} \frac{d\theta}{2\pi}
\cos^2(\theta-\alpha)\Delta^2(\theta),
\end{eqnarray}
and there arises an anisotropy as a function of $\alpha$ in the
nonlinear coefficient as $T\to T_c$, which is not there in pure
d-wave. Indeed, it can be sizable as shown in Fig.~\ref{fig6} (upper frame)
for different percentages of the s-wave component. For 15\% s-wave,
the anisotropy is 20\%. Note also that the anisotropy is twofold.
At low temperature, a d-wave order parameter already shows a 
fourfold anisotropy in $\beta(T)$ as a function of $\alpha$. If we
plot $\beta(T)$ normalized to $\Delta_0/24T$ as is shown in the lower
frame of Fig.~\ref{fig6}, we see that the fourfold d-wave anisotropy
shifts to twofold for d$+$s-wave and also exhibits a shift in the
position of the maxima (indicating a shift in the position of the nodes).
The magnitude of the anisotropy, which was a factor of 2 for d-wave, is
now even greater in  d$+$s-wave. The analytic formula for this anisotropy,
assuming $c<a$,
is 
\begin{eqnarray}
\beta(T)&=&\frac{\Delta_0}{24T}
\biggl(\biggl[1-\frac{c}{a}\cos(2\alpha)\biggr]^2+\biggl[1-\biggl(\frac{c}{a}\biggr)^2\biggr]\sin^2(2\alpha)\biggr)\nonumber\\
&&\times\frac{1}{\sqrt{1-(c/a)^2}}.
\end{eqnarray}
Thus, there should be observable signatures of an s-wave component should it
be possible to do angle-resolved measurements of intrinsic nonlinear current.

\section{Conclusions}

In summary, we have calculated the intrinsic nonlinear current of
a d-wave superconductors, including the effect of the superfluid current
on the order parameter. We find that the low temperature $1/T$ behavior
of the nonlinear coefficient, proposed as a test for d-wave symmetry 
by Dahm and Scalapino, remains unchanged. However, at temperatures
above about $0.2T_c$, the approximation of taking the gap to be
$q_s$-independent fails and large corrections are found. Indeed, at
temperatures of order $0.75T_c$, there is a twofold increase in the nonlinear
coefficient over that obtained with a $q_s$-independent gap. This could have
implications for the technological application of the cuprate materials
as passive microwave devices for the communication industry. We also find
that these corrections remain large in the presence of impurity scattering.

A re-examination of the comparison of theory and
data from the intermodulation power with this approximation removed finds
that strong coupling  effects would still be required to obtain a fit over
the entire temperature range. However, this
requires calculations of such sophistication that they are
beyond the scope of this paper. The signature in support of
d-wave symmetry remains clear,
regardless. More importantly, a recent analysis of data from 
several films assumes that the explanation of the variation in the
nonlinear coefficient and penetration depth results from impurity scattering.
As the data is taken at high temperature, where the decay of the gap by the
current is significant, we find upon reevaluation of the theory that
the relationship between these two quantities can vary markedly and a 
different slope is obtained along with a different overall scale factor.
Unfortunately, we must point out that, while a possible fit to the data
still remains, the basic assumption that impurity scattering is the source
of the changes seen in the data implies a large variation in the $T_c$
of the different samples, which is not seen experimentally. Further
experiments along these lines might help to resolve this issue more
completely.

Finally, due to a recent experimental observation of possible d$+$s-wave
gap symmetry in YBCO measured by angled-resolved electron tunneling,
we have examined the nonlinear coefficient at both low $T$ and $T\to T_c$
to determine signatures of the s-wave component. Indeed, at $T\to T_c$,
the isotropic behavior found for pure s- and pure d-wave is lost and
a large anisotropy develops for d$+$s-wave as a 
function of the direction of the current in the plane.
At low temperatures, the 
$1/T$ dependence remains but the fourfold anistropy of the coefficient
is altered to twofold and the magnitude is modified. Therefore, should it
become possible to do experiments where the direction of the current can
be varied with respect to the antinodes or nodes, then we predict that
signatures of a d$+$s-wave order parameter would be observable.

\begin{acknowledgments}
This work has been supported by NSERC (EJN and JPC) 
and the CIAR (JPC). We thank Prof. D.J. Scalapino for
interest and helpful discussions. 
\end{acknowledgments}


\begin{thebibliography}{99}

\bibitem{dswave} H.J.H. Smilde, A.A. Golubov, Ariando, G. Rijnders,
J.M. Dekkers, S. Harkema, D.H.A. Blank, H. Rogalla, and H. Hilgenkamp,
Phys. Rev. Lett. {\bf 95}, 257001 (2005).

\bibitem{specheat} Initial theoretical proposal by
I. Vekhter, P.J. Hirschfeld, J.P. Carbotte, E.J. Nicol, Phys. Rev. B, Rapid Comm. {\bf 59}, R9023 (1999) and
 I.Vekhter, P.J. Hirschfeld, and E.J. Nicol, 
Phys. Rev. B {\bf 64}, 064513 (2001) with experimental examples by
T. Park, M.B. Salamon, E.M. Choi, H.J. Kim, and
S.-I. Lee, Phys. Rev. Lett. {\bf 90},
177001 (2003); T. Park, E.E.M. Chia, M.B. Salamon, E.D. Bauer,
I. Vekhter, J.D. Thompson, E.M. Choi, H.J. Kim, S.-I. Lee,
and P.C. Canfield, Phys. Rev. Lett. {\bf 92},
237002 (2004); K. Deguchi, Z.Q. Mao, H. Yaguchi, and Y. Maeno,
Phys. Rev. Lett. {\bf 92}, 047002 (2004);
H. Aoki, T. Sakakibara, H. Shishido, R. Settai, Y. Onuki,
P. Miranovi\'c, and K. Machida, J. Phys.: Condens. Matter {\bf 16}
L13 (2004).

\bibitem{yip} S.K. Yip and J.A. Sauls, Phys. Rev. Lett. {\bf 69},
2264 (1992)

\bibitem{xu} D. Xu, S.K. Yip, and J.A. Sauls, Phys. Rev. B {\bf 51},
16233 (1995).


\bibitem{saulsexpt} A survey  of the experiments 
has been given in Ref.~\cite{dahmdwave}. The paper by
Klaus Halterman, Oriol T. Valls, and
Igor \v Zuti\'c, Phys. Rev. B
{\bf 63} 180405(R) (2001) provides
a more recent experimental analysis that  claims to observe the NLME.


\bibitem{dahmdwave} T. Dahm and D.J. Scalapino, Phys. Rev. B {\bf 60}, 13125 
(1999).

\bibitem{dahmapj} T. Dahm and D.J. Scalapino, J. Appl. Phys. {\bf 81}, 2002
(1997).

\bibitem{benz} G. Benz, S. W\"unsch, T.A. Scherer, M. Neuhaus, W. Jutzi,
Physica C {\bf 356}, 122 (2001); 

\bibitem{oates} D.E. Oates, S.H. Park, and G. Koren, Phys. Rev. Lett. 
{\bf 93}, 197001 (2004); 

\bibitem{leong} K.T. Leong, J.C. Booth, and S.A. Schima, IEEE Trans.
Appl. Supercond. {\bf 15}, 3608 (2005).
{\bf 93}, 197001 (2004); 

\bibitem{agassi} D. Agassi and D.E. Oates, Phys. Rev. B {\bf 72}, 014538 (2005).

\bibitem{mrli} M.-R. Li, P.J. Hirschfeld, and P. W\"olfle, Phys. Rev. Lett.
{\bf 81} 5640 (1998).

\bibitem{andersen} B.M. Andersen, J.C. Booth, and P.J. Hirschfeld,
cond-mat/0510015


\bibitem{salkola} M.I. Salkola and D.J. Scalapino, Appl. Phys. Lett. {\bf 86}, 
112509 
(2005).

\bibitem{dahmmg} T. Dahm and D.J. Scalapino, Appl. Phys. Lett. {\bf 85}, 4436 
(2004).

\bibitem{nicolmg} E.J. Nicol, J.P. Carbotte, and D.J. Scalapino, 
Phys. Rev. B {\bf 73} in press (2006).

\bibitem{maki} K. Maki, in Superconductivity, edited by
R.D. Parks (Dekker, New York, 1969), p.1035.

\bibitem{nicoljc} E.J. Nicol and J.P. Carbotte, Phys. Rev. B {\bf 43},
10210 (1991).

\bibitem{mgjcnicol} E.J. Nicol and J.P. Carbotte, Phys. Rev. B {\bf 72},
014520 (2005).

\bibitem{comment} Many works give a real axis form for the
current which is in terms of a quasiparticle piece and a 
condensate piece, such as found, for example, in Refs.~\cite{dahmdwave,xu}.
Note that the order of the energy integration and
the Matsubara sum is important in constructing this quantity from
the original definition of the current\cite{xu}. If the energy integration
is done first, Eq.~(\ref{eq:js}) results and the expression contains
both quasiparticle and condensate terms.  

\bibitem{arberg} P. Arberg and J.P. Carbotte, Phys. Rev. B {\bf 50},
3250 (1994).

\bibitem{asymp} J.P. Carbotte and C. Jiang, Phys. Rev. B {\bf 48} R4231
(1993).

\bibitem{ting} D. Zhang, C.S. Ting, and C.-R. Hu, Phys. Rev. B {\bf 70},
172508 (2004).

\bibitem{hykee} I. Khavkine, H.Y. Kee, and K. Maki, Phys. Rev. B
{\bf 70}, 184521 (2004).

\bibitem{spin} E. Schachinger, J.P. Carbotte, and D.N.Basov,
Euro. Phys. Lett. {\bf 54}, 380 (2001); J.P. Carbotte, E. Schachinger,
and D.N. Basov, Nature {\bf 401} 354 (1999).

\bibitem{quinlan} Our variation of $T_c$ and penetration depth with
impurity scattering is in agreement 
 with the results found by P.J. Hirschfeld, W.O. Putikka,
and D.J. Scalapino, Phys. Rev. B {\bf 50} 10250 (1994).

\end{thebibliography}
\end{document}